\documentstyle[12pt]{article}
\begin{document}

\begin{center}
{\Large \bf Contradictory implications of \\the nonadditive entropy} \\
\vskip .5cm
K. Ropotenko\\
\centerline{\it State Service for Special Communication and}
\centerline{\it Information Protection of Ukraine} \centerline{\it
5/7 Patorzhynskoho str., Kyiv, 01034, Ukraine}
\bigskip
\verb"ropotenko@ukr.net"

\end{center}
\bigskip\bigskip

\begin{abstract}
It is shown that the concept of nonadditive black hole entropy leads
to the contradictory implications in the framework of statistical
thermodynamics. In particular, a black hole with the nonadditive
entropy cannot be in thermal equilibrium with ordinary matter.
Moreover, such black holes are mutually exclusive, i.e. they cannot
compose a single system.
\end{abstract}
\bigskip\bigskip

According to statistical mechanics the entropy of a thermodynamical
system is the logarithm of the number of microstates accessible to
it $\Gamma$, that is,
\begin{equation}
\label{eq1} S = \ln \Gamma.
\end{equation}
The black hole is a thermodynamical system with the
Bekenstein-Hawking entropy
\begin{equation}
\label{eq2} S_{\rm BH}=\frac{A}{4 l_P^{2}},
\end{equation}
where $A$ is the area of the event horizon. A central problem in
black hole physics is to express the Bekenstein-Hawking entropy in
terms of the microstates (\ref{eq1}). The essential reason for
taking the logarithm in (\ref{eq1}) is to make the entropy of a
conventional system an \emph{additive} quantity, for the statistical
independent systems. In the literature there are however some doubts
about additivity of black holes \cite{kab}. The point is that the
Bekenstein-Hawking entropy is not a homogeneous first order function
of the black hole energy. Moreover, we cannot divide a black hole
into two independent subsystems by a partition as an ideal gas in a
box (the area theorem). And the black hole constituents cannot be
extracted from a black hole. Therefore the black hole cannot be
thought as made up of any constituent subsystems each of them
endowed with its own independent thermodynamics; we have to consider
a single black hole as a whole system.

In accord with this ideas I suggested in \cite{ro1} that the
statistical entropy of a black hole is not the logarithm of the
number of microstates (\ref{eq1}) but is proportional to this number
\begin{equation}
\label{eq3} S_{bh}=2\pi \Gamma,
\end{equation}
where
\begin{equation}
\label{eq4} \Gamma=\frac{A}{8 \pi l_P^{2}}.
\end{equation}
This means that the black hole is a \emph{nonadditive} system.

In this note I argue that a black hole with the nonadditive entropy
(\ref{eq3}) cannot be in thermal equilibrium with ordinary matter.
Moreover such black holes are mutually exclusive, i.e. they cannot
compose a single system. I show that the concept of the nonadditive
entropy (\ref{eq3}) leads to the contradictory conclusions in the
framework of the standard thermodynamics. This conclusion can be
relevant for nonextensive statistical mechanics \cite{tsal}.

The argument is simple and goes as follows. Consider a black hole
and ordinary matter in thermal equilibrium with each other, forming
an isolated system \cite{haw}. For simplicity I consider a
Schwarzschild black hole. Denote the energy and entropy of the black
hole as $E_{bh}$ and $S_{bh}$, and the energy and entropy of
ordinary matter as $E_{mat}$ and $S_{mat}$. Assume that 1) the
principle of maximum entropy is valid and 2) the energies and
entropies are additive for weakly coupled subsystems:
$E=E_{bh}+E_{mat}$ and $S=S_{bh}+S_{mat}$. Then the entropy $S$ of
the system has its maximum value for a given energy $E$ of the
system. Since the total energy is fixed, $S$ is really a function of
one independent variable, say $E_{bh}$, and the necessary condition
for a maximum may be written
\begin{equation}
\label{eq5} \frac{d S}{d
E_{bh}}=\frac{dS_{bh}}{dE_{bh}}+\frac{dS_{mat}}{dE_{mat}}\frac{dE_{mat}}{dE_{bh}}=0
\end{equation}
or
\begin{equation}
\label{eq6} \frac{dS_{bh}}{dE_{bh}}=\frac{dS_{mat}}{dE_{mat}}.
\end{equation}
I also assume that 3) the entropy (\ref{eq3}) satisfies the same
thermodynamical relations as the Bekenstein-Hawking entropy, in
particular $dS_{bh}/dE_{bh}=1/T_{\rm H}$, $T_{\rm H}$ being the
Hawking temperature, $T_{\rm H}=1/(8 \pi E_{bh})$. Then the
condition for equilibrium is
\begin{equation}
\label{eq7} T_{\rm H}=T_{mat}.
\end{equation}
This result is obtained from purely thermodynamical reasoning,
without any statistical assumptions about the form of the entropy.
Obviously it has the same form as for the conventional subsystems.
Consider now the same condition from a statistical mechanical point
of view. Denote the number of microstates accessible to the black
hole and ordinary matter as $\Gamma_{bh}$ and $\Gamma_{mat}$. Assume
now that 1) the black hole and ordinary matter  are statistically
independent so that the number of microstates accessible to the
whole system is
\begin{equation}
\label{eq8} \Gamma=\Gamma_{bh} \Gamma_{mat},
\end{equation}
and 2) thermal equilibrium is realized by the greatest number of
microstates, so that we can maximize this expression with respect to
$E_{bh}$ by writing
\begin{equation}
\label{eq9} \frac{d\Gamma}{dE_{bh}}=
\Gamma_{mat}\frac{d\Gamma_{bh}}{dE_{bh}}+
\Gamma_{bh}\frac{d\Gamma_{mat}}{dE_{mat}}\frac{dE_{mat}}{dE_{bh}}=0
\end{equation}
or
\begin{equation}
\label{eq10}\frac{1}{\Gamma_{bh}}\frac{d\Gamma_{bh}}{dE_{bh}}=
\frac{1}{\Gamma_{mat}}\frac{d\Gamma_{mat}}{dE_{mat}}.
\end{equation}
For the entropy of ordinary mater we have the standard formula with
the logarithm (\ref{eq1}). But for the black hole we have
(\ref{eq3}) and
\begin{equation}
\label{eq11}\frac{d\Gamma_{bh}}{dE_{bh}}=
\frac{1}{2\pi}\frac{dS_{bh}}{dE_{bh}}=\frac{1}{2\pi}\frac{1}{T_{\rm
H}}.
\end{equation}
So
\begin{equation}
\label{eq12}\frac{1}{2\pi \Gamma_{bh}}\frac{1}{T_{\rm
H}}=\frac{1}{T_{mat}}
\end{equation}
or
\begin{equation}
\label{eq13}\frac{1}{S_{bh}}\frac{1}{T_{\rm H}}=\frac{1}{T_{mat}}
\end{equation}
But $E_{bh}=2T_{\rm H}S_{bh}$. Then
\begin{equation}
\label{eq14} \frac{E_{bh}}{2}=T_{mat}
\end{equation}
or
\begin{equation}
\label{eq15} \frac{1}{16\pi T_{\rm H}}=T_{mat}
\end{equation}
But this violates the zeroth law of thermodynamics for systems in
thermal equilibrium. It is obvious that we cannot redefine the black
hole temperature by simply setting $T_{bh}\equiv 1/(16\pi T_{\rm
H})$. Moreover, this relation does not agree with (\ref{eq7}). Thus
the concept of thermal equilibrium cannot be formulated for black
holes with the nonadditive entropy (\ref{eq3}).

Note that this conclusion is valid not only for the formula
(\ref{eq3}) but also for the standard formula (\ref{eq1}) if
$\Gamma$ equals (\ref{eq4}). Moreover, the formula $S_{bh}=\ln (A/8
\pi l_P^{2})$ contradicts the second law of black hole
thermodynamics \cite{gour}, \cite{ro1}.

Consider now a system of black holes. Suppose, that two black holes
are far apart and their interaction is negligible, so that they can
be viewed as statistically independent. Let $S_{1(2)}=2\pi
\Gamma_{1(2)}$ and $\Gamma_{1(2)}$ be the entropy and degeneracy of
the first (second) black hole, respectively. Then the number of
states for the combined system is
\begin{equation}
\label{eq16} \Gamma=\Gamma_1\Gamma_2.
\end{equation}
What is the entropy of the system? Obviously, we cannot write the
total entropy as $S=2\pi (\Gamma_1 \Gamma_2)$ because our system is
not a single black hole. It seems that we would take the logarithm
of $\Gamma$: $\ln \Gamma = \ln \Gamma_1 + \ln \Gamma_2$. But in this
case, as mentioned above, we cannot interpret $\ln \Gamma_{1(2)}$ as
the entropy of the first (second) black hole. Despite this failure
the laws of thermodynamics are still valid, so we may define the
total entropy as
\begin{equation}
\label{eq17} S=S_1+S_2=2\pi \Gamma_1+2\pi \Gamma_2=2\pi
(\Gamma_1+\Gamma_2),
\end{equation}
whence
\begin{equation}
\label{eq18} \Gamma_{total}=\Gamma_1+\Gamma_2.
\end{equation}
This means that these two black holes are mutually exclusive, i.e.
no two black holes can be simultaneously in a single system. But
this does not agree with (\ref{eq16}). We can extend this conclusion
to an arbitrary number of black holes. Note that additivity of
entropies (\ref{eq17}) is valid even when the subsystems cannot be
considered independent and interact strongly among themselves; it is
a consequence of the additivity of actions in a path integral
approach to statistical thermodynamics \cite{mart}.

Thus the concept of the nonadditive entropy leads to the
contradictory conclusions in the framework of the standard
thermodynamics.

In conclusion, the following point may be noted. In deriving
(\ref{eq3}) in \cite{ro1}, I used the concept of the internal
(Euclidean) angular momentum of a black hole $L_z=A/8\pi$. Although
its identification with the number of microstates (\ref{eq4}) is not
correct, this concept is well established. In \cite{ro2}, by
following the approach used by Susskind \cite{sus} to derive the
Rindler energy, I obtained quantization of the black hole area from
the commutation relation and quantization condition for $L_z$. But
$L_z$ can be defined in more simple way from the Bunster-Carlip
equation \cite{car}
\begin{equation}
\label{eq20} \frac{\hbar}{i}\frac{\partial \psi}{\partial
\Theta}-\frac{A}{8\pi}\psi=0,
\end{equation}
where $\Theta$ is the lapse of the hyperbolic angle at the horizon.
Analytically continuing  $\Theta$ and $A/8\pi$ to the real values of
$\Theta_{\rm E}=i\Theta$ and $(A/8\pi)_{\rm E}=-i(A/8\pi)$ we obtain
\begin{equation}
\label{eq21} -i \hbar\frac{\partial \psi}{\partial \Theta_{\rm
E}}-\left(\frac{A}{8\pi}\right)_{\rm E}\psi=0,
\end{equation}
As a result, $(A/8\pi)_{\rm E}$ and $\Theta_{\rm E}$ become
conjugate. This means that the area is the operator-valued quantity,
the angular momentum. Indeed, in the semiclassical approximation
\begin{equation}
\label{eq22} \psi=a\exp \left(\frac{i}{\hbar}I\right),
\end{equation}
where $I$ is the action of a black hole. Substituting this in
(\ref{eq20}) we obtain
\begin{equation}
\label{eq23} \frac{\partial I}{\partial
\Theta}\psi=\frac{A}{8\pi}\psi;
\end{equation}
the slowly varying amplitude $a$ need not be differentiated. Under
Euclidean continuation $\Theta_{\rm E}=i\Theta$ and $(A/8\pi)_{\rm
E}=-i(A/8\pi)$,
\begin{equation}
\label{eq24} \frac{\partial I}{\partial \Theta_{\rm E}\psi
}=\left(\frac{A}{8\pi}\right)_{\rm E}\psi.
\end{equation}
The derivative $\partial I/\partial \Theta_{\rm E}$ is just a
generalized momentum corresponding to the angle of rotation about
one of the axes (say, the $z^{\rm{th}}$) for a mechanical system.
Therefore the operator $(A/8\pi)_{\rm E}$ is what corresponds in
quantum mechanics to the $z$ component of angular momentum
$\hat{L}_z$. Medved \cite{med} found it immediately from the
Bunster-Carlip action \cite{car}.

\end{document}